\documentclass[aps,floatfix,twocolumn]{revtex4}
\usepackage{graphics,epsfig}
\usepackage{graphicx}
\usepackage{amsmath,amssymb}
\usepackage{dcolumn}

\begin{document}

\title{Generalized Swiss-cheese cosmologies: Mass scales}
\author{C\'{e}dric Grenon\footnote{cgrenon@astro.queensu.ca} and Kayll Lake\footnote{lake@astro.queensu.ca}}
\affiliation{Department of Physics, Queen's University, Kingston,
Ontario, Canada, K7L 3N6 }
\date{\today}

\begin{abstract}
We generalize the Swiss-cheese cosmologies so as to include nonzero linear momenta of the associated boundary surfaces. The evolution of mass scales in these generalized cosmologies is studied for a variety of models for the background without having to specify any details within the local inhomogeneities. We find that the final effective gravitational mass and size of the evolving inhomogeneities depends on their linear momenta but these properties are essentially unaffected by the details of the background model.
\end{abstract}
\maketitle
\section{Introduction}
The Swiss-cheese models give us (noninteracting) inhomogeneities in a cosmological setting that are exact solutions to Einstein's equations. As a result, the models have become a standard construction \cite{forms} and are very widely studied \cite{2}. The classical Einstein-Straus vacuole model, which requires a comoving boundary surface, is unstable. Many subsequent Swiss-cheese models have also assumed that the associated boundary surfaces remain comoving, but it is well known that this need not be the case. Some general studies of this issue go back many years \cite{lake}. Some specific examples of noncomoving boundary surfaces include the study of Vaidya-type inhomogeneities \cite{fayos} and the evolution of ``density waves" in the Lema\^{\i}tre-Tolman model \cite{lt}. Here we examine the role of the linear momentum of a boundary surface in a Robertson-Walker background. Within the context of these generalized models, we study the evolution of the effective gravitational mass and size of the inhomogeneities for a variety of well-known background models. We find that whereas the momentum plays a central role, the details of the background model are relatively unimportant.
\section{Model Construction}
The theory of hypersurfaces in spacetime is well established (see, for example, \cite{poisson}) and we do not reproduce all the necessary machinery here. Rather, we go directly to the essential ingredients of the model. The model consists of randomly distributed nonintersecting spherical boundary surfaces $\Sigma$ in a Robertson-Walker background with each particle on $\Sigma$ executing radial timelike geodesic motion.
\subsection{Junction Conditions}
To establish notation, let us write the metrics of the spacetimes $\mathcal{V}_{\pm}$ in the form \cite{notation}
\begin{equation}\label{spacetimes}
    ds^2_{\pm}=ds^2_{\Gamma_{\pm}}(x^1_{\pm},x^2_{\pm})+R^2(x^1_{\pm},x^2_{\pm})d\Omega^2\,,
\end{equation}
where the signature of the two-surfaces $\Gamma_{\pm}$ is zero and $d\Omega^2$ is the metric of a unit two-sphere, which we write in the usual form $d\theta^2+\sin^2(\theta)d\phi^2$. We need consider only one boundary surface. The metric on $\Sigma$ can be written in the form
\begin{equation}\label{sigma}
    ds^2_{\Sigma}=R^2(\tau)d\Omega^2-d\tau^2\,,
\end{equation}
where $\tau$ is the proper time on $\Sigma$. Since $\Sigma$ is, by assumption, geodesic, there is only one nonvanishing independent component of the extrinsic curvature $K_{\alpha\beta}$. This can be written in the form (e.g. \cite{lake})
\begin{equation}\label{extrinsic}
    K^2_{\theta \theta_{\pm}}=R^2\left(\dot{R}^2+1-\frac{2\mathcal{M}_{\pm}}{R}\right)\,,
\end{equation}
where $^{.} \equiv d/d\tau$ and $\mathcal{M}_{\pm}$ are the effective gravitational masses of $\mathcal{V}_{\pm}$. The invariant
properties of $\mathcal{M}$ were first explored by Hernandez and Misner
\cite{hm} who wrote the function in the form
\begin{equation}
\mathcal{M}= \frac{R^{3}}{2}\mathcal{R}_{\theta \phi}^{\; \; \;\; \theta
\phi} , \label{mass}
\end{equation}
where $\mathcal{R}$ is the Riemann tensor of $\mathcal{V}$. See
also \cite{cm, pe, z, h} for further
discussion \cite{covariant}. From (\ref{extrinsic}), and the continuity conditions, we arrive at the central condition of the model:
\begin{equation}\label{massc}
    \mathcal{M}_{-}=\mathcal{M}_{+},
\end{equation}
a statement which is independent of the coordinates $(x^1_{\pm},x^2_{\pm})$.
\subsection{Background Geodesics}
The Robertson-Walker geometry (excluding the Einstein static subcase) has a Killing algebra of dimension 6 (3 translations and 3 rotations). Since $\Sigma$ is in radial motion, we are interested in the constants of motion generated by translational invariance. Write the background in the form
\begin{equation}\label{rw}
    ds^2_{+}=a^2(t)\left(\frac{dr^2}{\epsilon^2(r)}+r^2d\Omega^2\right)-dt^2\,,
\end{equation}
where $\epsilon(r)\equiv \sqrt{1-kr^2}$ with $k=\pm 1,0$. The translational Killing vectors are then given by
\begin{equation}\label{one}
    \xi_{1}^{\alpha}=\epsilon\left(\cos(\theta) \partial_{r}-\frac{\sin(\theta)}{r} \partial_{\theta}\right),
\end{equation}

\begin{equation}\label{two}
    \xi_{2}^{\alpha}=\epsilon\left[\sin(\theta) \sin(\phi) \partial_{r}+\frac{1}{r}\left(\cos(\theta) \sin(\phi) \partial_{\theta}+\frac{\cos(\phi)}{\sin(\theta)} \partial_{\phi}\right)\right],
\end{equation}
and
\begin{equation}\label{three}
    \xi_{3}^{\alpha}=\epsilon\left[\sin(\theta) \cos(\phi) \partial_{r}+\frac{1}{r}\left(\cos(\theta) \cos(\phi) \partial_{\theta}-\frac{\sin(\phi)}{\sin(\theta)} \partial_{\phi}\right)\right].
\end{equation}
Consider a radial geodesic with tangent $u^{\alpha}=\dot{r} \partial_{r}+\dot{t} \partial_{t}$ and define the constants $\mathcal{C}_{n}\equiv\xi_{n}^{\alpha} u_{\alpha}$ and $\mathcal{D}^2 \equiv \sum_{n}\mathcal{C}^2_{n}$. It follows directly from (\ref{one})--(\ref{three}) that
\begin{equation}\label{rdot}
    \dot{r}^2=\frac{\mathcal{D}^2 \epsilon^2}{a^4},
\end{equation}
so that from the timelike condition $u^{\alpha}u_{\alpha}=-1$ we have
\begin{equation}\label{tdot}
    \dot{t}^2=1+\frac{\mathcal{D}^2}{a^2}.
\end{equation}
Equations (\ref{rdot}) and (\ref{tdot}) should be well known \cite{ellis}. Our purpose here is to explain the physical meaning of $\mathcal{D}$. It is the total linear momentum of $\Sigma$ \cite{hobson}.
\subsection{Swiss-cheese}\label{Swiss-cheese}
The standard Swiss-cheese inhomogeneous cosmology takes $\mathcal{D}=0$. Of these models, the Einstein-Straus case, which sets $\mathcal{V}_{-}$ as vacuum, is the most well known. The model has seen very wide application \cite{forms}, but it is known to be unstable: to aspherical perturbations \cite{mars1} and to perturbations in the condition (\ref{massc}) which could lead to the development of surface layers \cite{krasinski}. A clear way to see the instability in this model at a primitive level is to look at the necessary conditions for a boundary surface:
\begin{equation}\label{jump}
    [G^{\alpha}_{\beta}n_{\alpha}u^{\beta}]=[G^{\alpha}_{\beta}n_{\alpha}n^{\beta}]=0,
\end{equation}
where $G^{\alpha}_{\beta}$ is the Einstein tensor, $[\Psi] \equiv (\Psi_{+}-\Psi_{-})|_{\Sigma}$, and $u^{\alpha}$ and $n^{\alpha}$ are the tangent and normal vectors to $\Sigma$ respectively. By assumption, $\mathcal{V}_{-}$ satisfies
\begin{equation}\label{vacuum}
    G^{\alpha}_{\beta}+\Lambda \delta^{\alpha}_{\beta}=0\,,
\end{equation}
where $\Lambda$ is the cosmological constant. As a result, the following two conditions must be satisfied in $\mathcal{V}_{+}$ if $\Sigma$ constitutes a boundary surface:
\begin{equation}\label{flux}
    G^{\alpha}_{\beta}n_{\alpha}u^{\beta}=0,
\end{equation}
and
\begin{equation}\label{pressure}
    G^{\alpha}_{\beta}n_{\alpha}n^{\beta}+\Lambda=0.
\end{equation}
From (\ref{rw}) and (\ref{flux}) we find $\dot{r}=0$ and from (\ref{rw}) and (\ref{pressure}) we find $8 \pi p=0$ where $p$ is the comoving isotropic pressure. As a result, in the Einstein-Straus model, $\Sigma$ must be exactly comoving and the background must be exactly dust. If either of these conditions do not hold then (\ref{massc}) is necessarily violated. Without resorting to surface layers, we can take the view that the culprit is the assumption that $\mathcal{V}_{-}$ is exactly vacuum. That is, (\ref{vacuum}) is too strong a condition to impose on $\mathcal{V}_{-}$. In other Swiss-cheese cosmologies, (\ref{vacuum}) is not imposed on $\mathcal{V}_{-}$. A widely used example is to assume that $\mathcal{V}_{-}$ is dust. However, typically $\Sigma$ is taken as comoving \emph{a priori} and so we arrive back at (\ref{flux}) and (\ref{pressure}) which, we would like to emphasize, are not necessary conditions for a boundary surface.
\subsection{Generalized Model}
In this paper the only condition imposed on $\mathcal{V}_{-}$ is (\ref{massc}) and we concentrate on the evolution of $\Sigma$ in  $\mathcal{V}_{+}$ which, for the sake of clarity, we assume is spatially flat. It follows from (\ref{rdot}) and (\ref{tdot}) that \cite{lake1}
\begin{equation}\label{rsigma}
    r_{_{\Sigma}}(z)=\int_{z}^{\infty}\sqrt{\frac{\mathcal{D}^2(1+x)^2}{1+\mathcal{D}^2(1+x)^2}}\frac{dx}{H(x)}\,,
\end{equation}
where we have set the initial conditions by $r_{_{\Sigma}}(t=0)=0$; the Hubble function $H$ is given, as usual, by $\frac{1}{a}\frac{da}{dt}$; and, without loss in generality, we have set $a_{_{0}}$ (today)$=1$. We have chosen as an independent variable $z=1/a-1$ and so for universes with a big bang that do not recollapse $-1<z<\infty$.
\section{Background Models}
In this work we explore mass scales associated with various models for the background specified by $H(z)$. The inhomogeneities considered here are governed by one free parameter, the total linear momentum $\mathcal{D}$. The first task at hand then is to establish a reasonable range of values for this parameter.
\subsection{The Range in $\mathcal{D}$}
\begin{figure}[ht]
\epsfig{file=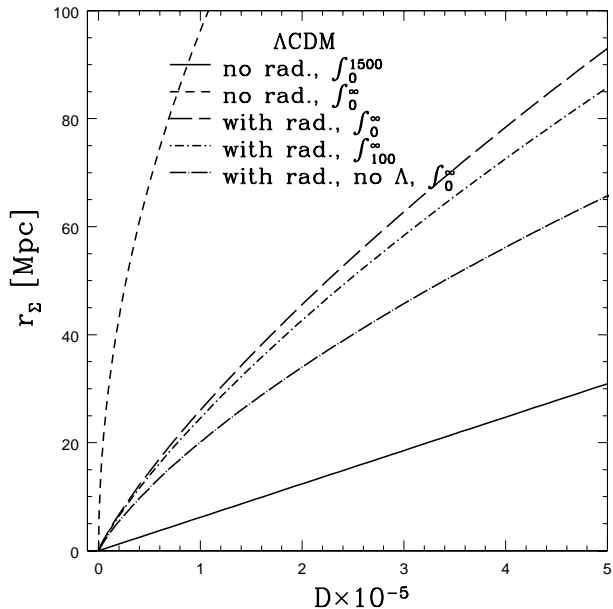,height=3.5in,width=3.5in,angle=0}
\caption{\label{fig1}The abscissa is the dimensionless fraction $\mathcal{D}/c \times 10^{-5}$ and the ordinate gives $r_{_{\Sigma}}(0)$ in units of Mpc. The top curve is the $\Lambda$CDM model. The middle curve is $\Lambda$CDM+noninteracting radiation (with the curve just below giving $r_{_{\Sigma}}(100)$ for comparison). The second to bottom curve is the noninteracting matter and radiation without $\Lambda$, and the bottom curve is the $\Lambda$CDM with integration upper limit going to $z=1500$. We have adopted the values $\Omega_{m_{0}}=0.27$, $h \equiv H_{0}/100=0.72$ and $T_{0}=2.73$.}\label{rsigma5}
\end{figure}

\begin{figure}[hb]
\epsfig{file=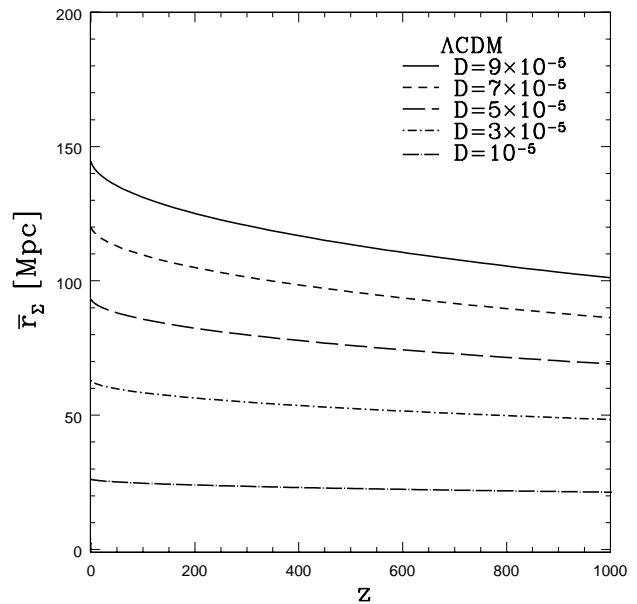,height=3.5in,angle=0}
\caption{Variation in the evolution of $\bar{r}_{_{\Sigma}}$ with redshift from variations in $\mathcal{D}$ for a $\Lambda$CDM background. We have used $h=0.72$ and $T_0=2.73$, and like all ``$\,\,\bar{ }\,\,$'' quantities we have used $\Omega_{m_{0}}=0.27$ and $\Omega_{R_0}=8\times10^{-5}$. These surfaces start with $\bar{r}_{_{\Sigma}}(z=\infty)=0$. The value of the linear momentum $\mathcal{D}$ determines the final size of these inhomogeneities.}\label{rsigmabar}
\end{figure}

To establish a reasonable range in $\mathcal{D}$ we evaluate $r_{_{\Sigma}}(0)$ from (\ref{rsigma}) for three standard models: $\Lambda$CDM, $\Lambda$CDM+noninteracting radiation (see the Appendix) and noninteracting matter and radiation without $\Lambda$. The results are shown in Fig.\ref{rsigma5}. The top curve is included so as to show the effect of ignoring the background radiation, and the two bottom curves are included so as to show the effect of ignoring $\Lambda$ and the effect of changing the integration upper limit to a redshift where radiation can be neglected, here $z=1500$. The middle curve shows the size of the hole for the $\Lambda$CDM+noninteracting radiation at earlier time $r_{_\Sigma}(100)$. The wide difference in size $r_{_{\Sigma}}$ between the top and bottom curves shows the effect of adding radiation on the evolution of the surfaces for a given value of $\mathcal{D}$. From the results of Fig.~\ref{rsigma5}, we have chosen to include radiation (see the Appendix) to all background models $H(z)$.
Throughout the paper, we define quantities associated with a flat ``standard'' $\Lambda$CDM model as `` $\bar{\,\,\,}$ '' when using the following values: for the matter energy density, $\Omega_{m_0}=0.27$, radiation energy density (including relativistic neutrinos), $\Omega_{R_0}\sim8\times10^{-5}$ and cosmological constant $\Omega_\Lambda$, regardless of the value of our free parameter $\mathcal{D}$. Using the $\Lambda$CDM background, Fig.~\ref{rsigmabar} shows the evolution of $\bar{r}_{_{\Sigma}}$ with total linear momentum $\mathcal{D}$ in the order of $\sim10^{-5}$. These inhomogeneities begin with $\bar{r}_{_{\Sigma}}(\infty)=0$ and grow mostly during the radiation epoch.
We are interested in ranges of the parameter $\mathcal{D}$ where surfaces $r_{_{\Sigma}}$ reach up to $\sim100$ Mpc. From Figs.~\ref{rsigma5} and \ref{rsigmabar}, this correspond to a range $10^{-5}\lesssim\mathcal{D}\lesssim5\times10^{-5}$. In subsequent figures we use $\mathcal{D}=5\times10^{-5}$ unless specified otherwise.
\subsection{Flat standard cosmological model ($\Lambda$CDM)}
\begin{figure}[ht]
\epsfig{file=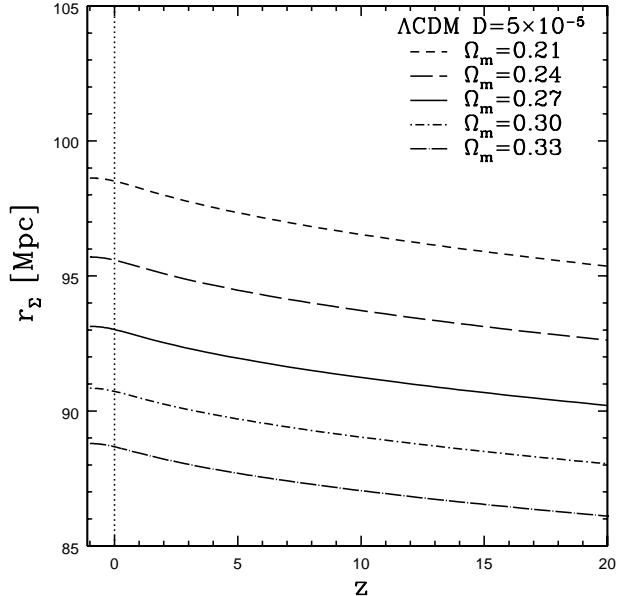,height=3.5in,angle=0}
\caption{Evolution of $r_{_{\Sigma}}$ with redshift for some values of the matter density $\Omega_{m}$. We have used a $\Lambda$CDM background with a total linear momentum of $\mathcal{D}=5\times10^{-5}$ and we have extrapolated the sizes of these inhomogeneities to the infinite future ($z=-1$). The dotted line shows the size at the present redshift ($z=0$). We have used values of $\Omega_m$ consistent with the 99.9\% confidence interval of \cite{Davis07} to show the effect of $\Omega_{m_0}$. For comparison, the latest WMAP5 \cite{Komatsu09} results give $\Omega_{m_0}=0.279\pm0.015$.}\label{H0rOm}
\end{figure}
%
\begin{figure}[hb!]
\epsfig{file=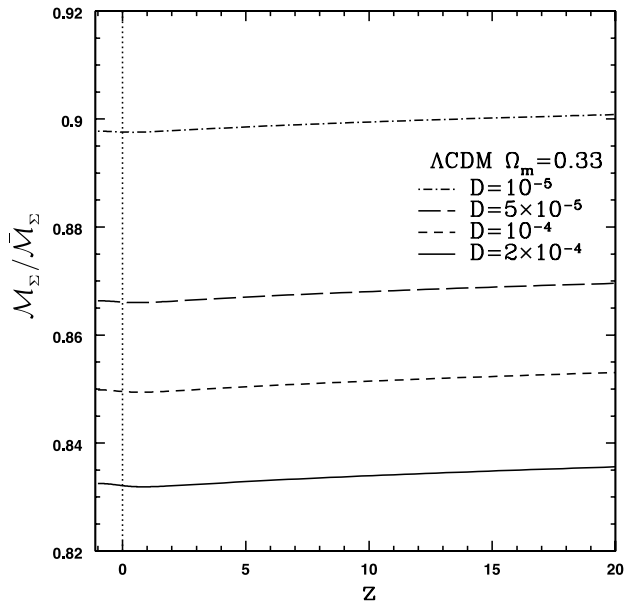,height=3.5in,angle=0}
\caption{Evolution of ratio $\mathcal{M}_{_{\Sigma}}/\bar{\mathcal{M}}_{_{\Sigma}}$ with redshift for some values of $\mathcal{D}=\bar{\mathcal{D}}$ where $\mathcal{M}$ is given by the $\Lambda$CDM with $\Omega_m=0.33$ and $\bar{\mathcal{M}}$ is our standard $\Lambda$CDM model.
}\label{MSMLCDM}
\end{figure}
%
\begin{figure}[ht!]
\epsfig{file=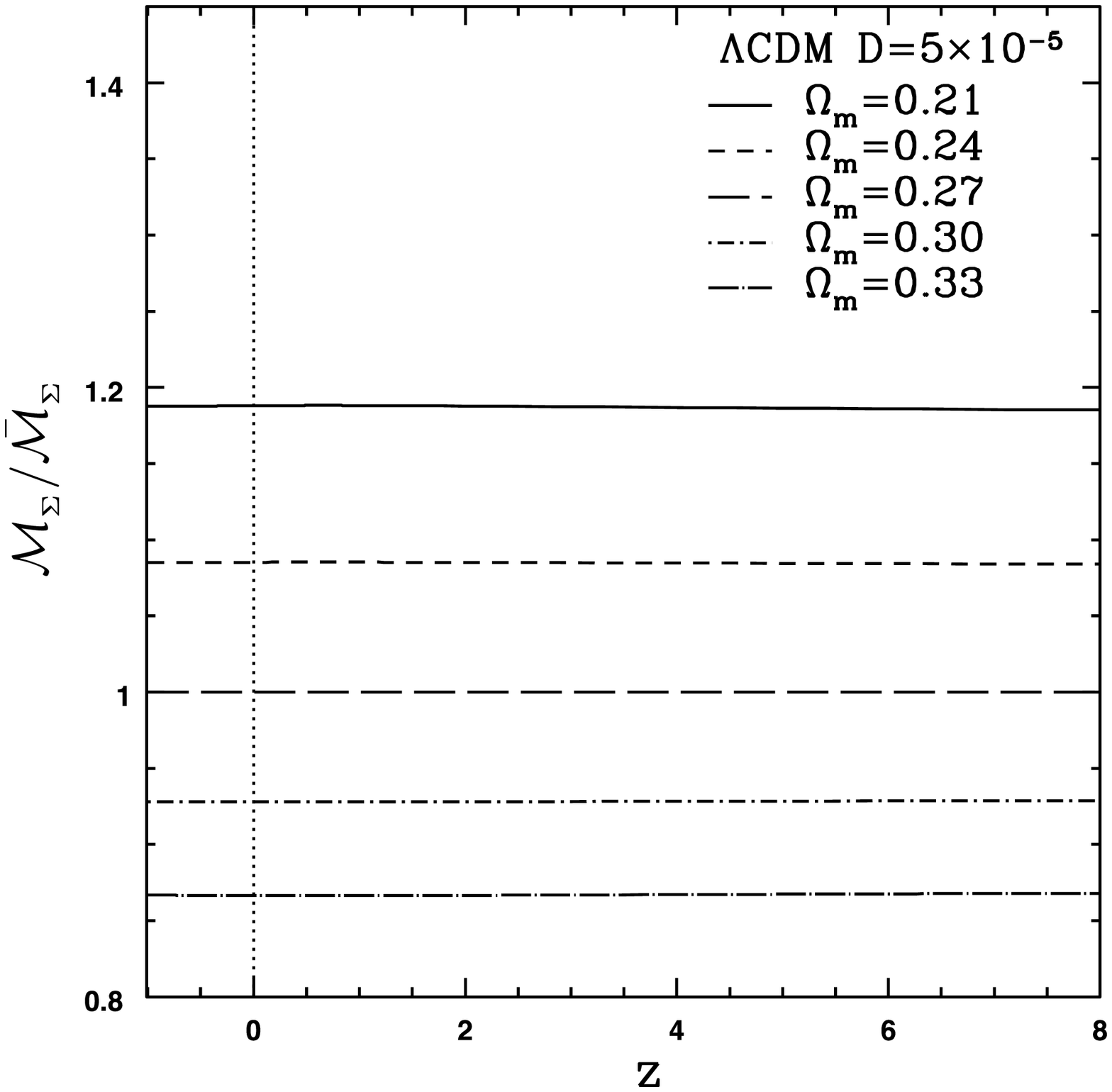,height=3.5in,angle=0}
\caption{Evolution of ratio $\mathcal{M}_{_{\Sigma}}/\bar{\mathcal{M}}_{_{\Sigma}}$ for $\Lambda$CDM with $\mathcal{D}=5\times10^{-5}$ for some values of $\Omega_m$.
}\label{MSMLCDM2}
\end{figure}
In the flat $\Lambda$CDM model, dark energy is the cosmological constant, and as usual we have
\begin{equation}
\left(\frac{H(z)}{H_0}\right)^2=\Omega_m(1+z)^3 +\Omega_\Lambda+\Omega_R(1+z)^4\,,
\end{equation}
where the $\Omega$ refer to current values. Note that we have explicitly included radiation density $\Omega_R$ (see the Appendix) and it is included implicitly in all the following background models. The influence of $\Omega_m$ on $r_{_{\Sigma}}$ is shown in Fig.~\ref{H0rOm}. The WMAP5 \cite{Komatsu09} value of $\Omega_{m_0}=0.279\pm0.015$ has limited effect on the final size compared to the value of the linear momentum in Fig.~\ref{rsigmabar}. To compare the different backgrounds, we can plot the mass ratio between the background model $\mathcal{M}_{_{\Sigma}}$ and the $\Lambda$CDM background $\bar{\mathcal{M}}_{_{\Sigma}}$. From (\ref{mass}) it follows that
\begin{equation}\label{massratio}
\frac{\mathcal{M}_{_{\Sigma}}}{\bar{\mathcal{M}}_{_{\Sigma}}}=\left(\frac{r_{_{\Sigma}}}{\bar{r}_{_{\Sigma}}}\right)^3\,.
\end{equation}
Throughout, for (\ref{massratio}), we use the same value for the linear momentum ($\mathcal{D}=\bar{\mathcal{D}}$). For the $\Lambda$CDM model, the mass ratio (\ref{massratio}) is shown in Fig.~\ref{MSMLCDM} where $\mathcal{M}$ is determined by $\Omega_{m_0}=0.33$ and different values of $\mathcal{D}$ are shown. In Fig.~\ref{MSMLCDM2} we plot the mass ratio (\ref{massratio}) for the $\Lambda$CDM model for the same values of $\Omega_{m_0}$ as Fig.~\ref{H0rOm}.\\

Define
\begin{equation}\label{deltam}
    \Delta\mathcal{M} \equiv \mathcal{M}_{_{\Sigma}}/\bar{\mathcal{M}}_{_{\Sigma}}-1.
\end{equation}
For $r_{_\Sigma} \sim \bar{r}_{_\Sigma}$ then $\Delta\mathcal{M} \sim 3\Delta r$, where $\Delta r=(r_{_\Sigma}/\bar{r}_{_\Sigma})-1$ is the size difference for a fixed $\mathcal{D}$. This can be gleaned from Figs.~\ref{rsigmabar}--\ref{MSMLCDM2}.
%
\subsection{Flat dark energy with constant equation of state}
\begin{figure}[ht!]
\epsfig{file=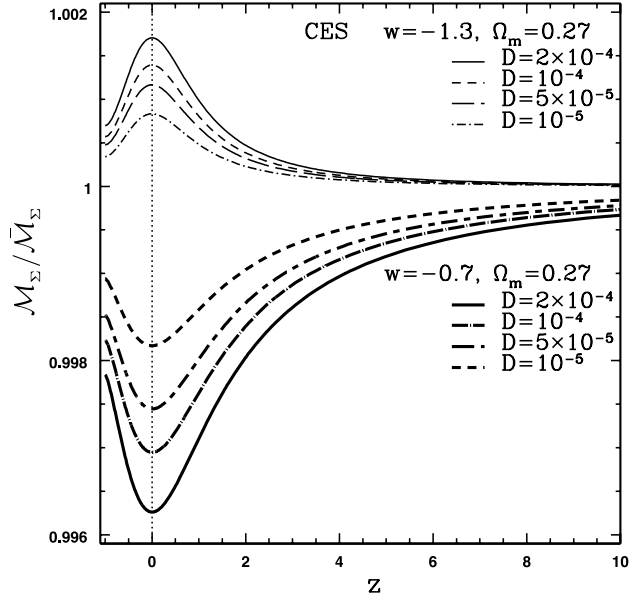,height=3.5in,angle=0}
\caption{The same as Fig.~\ref{MSMLCDM} where $\mathcal{M}$ now has the constant equation of state model (CES) for the background.}\label{CES}
\end{figure}

\begin{figure}[ht!]
\epsfig{file=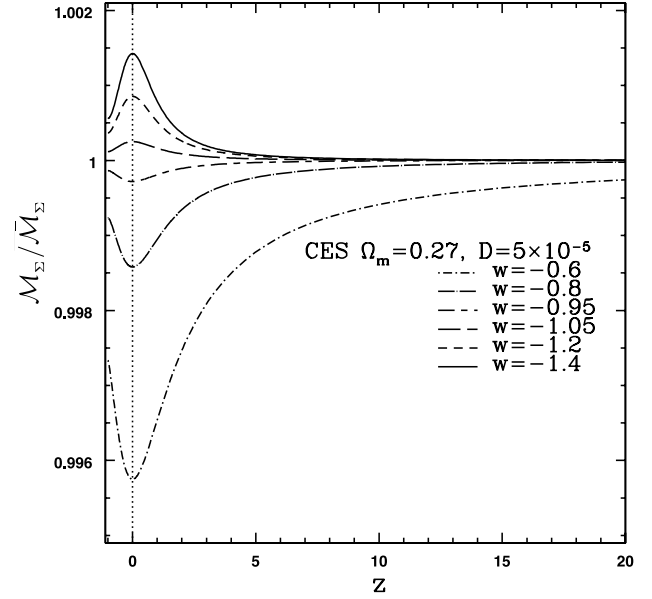,height=3.5in,angle=0}
\caption{Ratio $\mathcal{M}_{_{\Sigma}}/\bar{\mathcal{M}}_{_{\Sigma}}$ for $\mathcal{D}=5\times10^{-5}$ for different values of $w$ with $\Omega_m=0.27$ for the constant equation of state model.}\label{CES2}
\end{figure}

This model (which we designate by CES) uses a constant arbitrary value for the equation of state parameter $w \equiv p/\rho$ so that the Hubble function becomes
\begin{equation}
\left(\frac{H(z)}{H_0}\right)^2=\Omega_m(1+z)^3+\Omega_{de}(1+z)^{3(1+w)}.
\end{equation}
In Fig.\ref{CES} we show the ratio $\mathcal{M}_{_{\Sigma}}/\bar{\mathcal{M}}_{_{\Sigma}}$ for $w=-0.7$ and $w=-1.3$ corresponding to a 99.9\% confidence level \cite{Davis07}. The mass difference $\Delta\mathcal{M}$ between the two models is less than $2\%$ at maximum. The difference would be negligible using WMAP5 \cite{Komatsu09} where $w$ is constrained to $-0.97\pm0.06$ as can be seen in Fig.\ref{CES2} where we have plotted the mass ratio for different values of $w$ with $\Omega_m=0.27$.
%
%
\subsection{Flat dark energy with variable equation of state (VES)}
\begin{figure}[ht!]
\epsfig{file=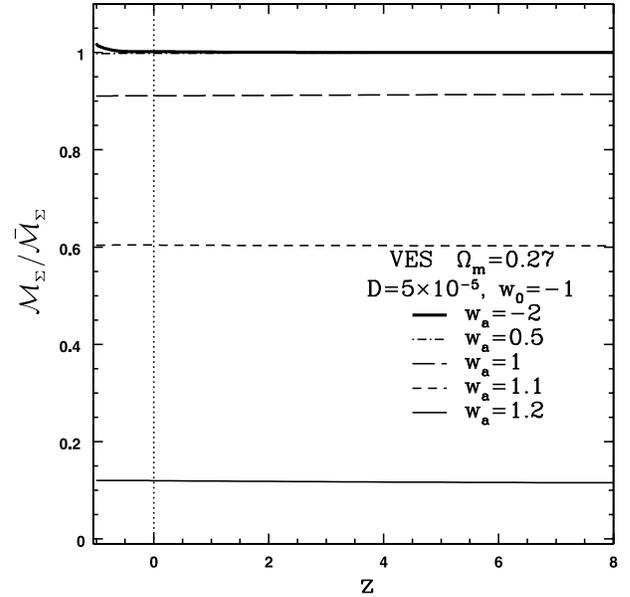,height=3.5in,angle=0}
\caption{The same as Fig.~\ref{CES2} with the variable equation of state (VES) model as background for $\mathcal{M}$. Only the region $w_a<1$ is of physical interest.}\label{VES}
\end{figure}

Allowing the equation of state to vary with time (VES), and using the parameterization \cite{ChevalierLinder}
\begin{equation}
w(z)=w_0+w_a\frac{z}{1+z}\,,
\end{equation}
we have
\begin{equation}\label{HVES}
\begin{split}
\left(\frac{H(z)}{H_0}\right)^2=&\Omega_m(1+z)^3 +\Omega_{de}(1+z)^{3(1+w_0+w_a)}\\
&\times e^{-3w_a\left(z/\left(1+z\right)\right)}\,.
\end{split}
\end{equation}
The situation is examined in Fig.\ref{VES}. Whereas the mass ratio decreases with increasing $w_a$, there is no evidence that $w_0+w_a> 0$ \cite{Kowalski08}.

%
\subsection{Flat DGP models}
The flat Dvali-Gabadadze-Porrati DGP model \cite{DGP00} (see also \cite{maartens}) is a one parameter model from brane theory where $\Omega_r=1/(4r^2H_0^2)$ is the dimensional parameter determined by the scale length $r$ which governs the transition from 4D to 5D behavior. For this model, the Hubble parameter is given by
\begin{equation}
\left(\frac{H(z)}{H_0}\right)^2= \left(\sqrt{\Omega_m(1+z)^3+\Omega_r}+\sqrt{\Omega_r}\right)^2\,,
\end{equation}
\noindent where $\Omega_m=1-2\sqrt{\Omega_r}$.
\begin{figure}[ht!]
\epsfig{file=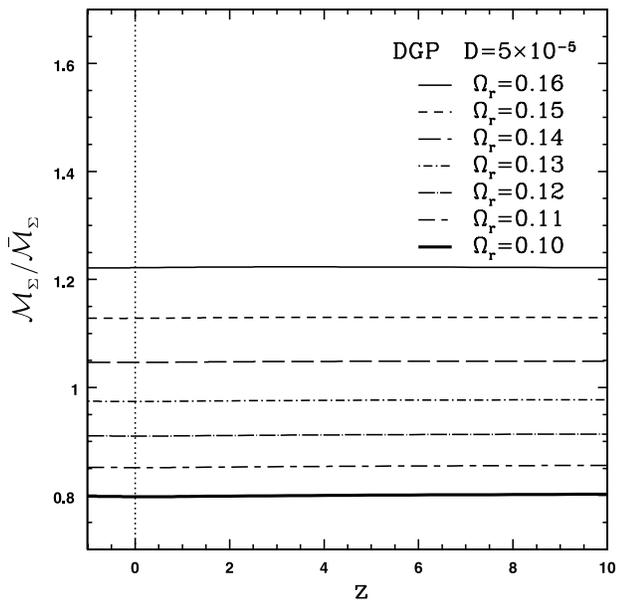,height=3.5in,angle=0}
\caption{Same as Fig.~\ref{CES2}; $\mathcal{M}$ is given for the Dvali-Gabadadze-Porrati (DGP) model.}
\label{DGP}
\end{figure}
The 99.9\% confidence level of \cite{Davis07} was used for the values of $\Omega_r$ in order to show the mass ratio (\ref{massratio}) for the DGP model as given in Fig.~\ref{DGP}. Subsequent models showing the same type of behavior as Fig.~\ref{DGP} over the redshift range $-1<z<20$ will be summarize in Table~\ref{table1}.
\begin{table*}
\caption{Model properties \label{table1}}
\begin{ruledtabular}
\begin{tabular}{ccccccc}
Model& $\Omega_m$ &$\Omega_{de}$& $1^{\text{st}}$ parameter& $2^{\text{nd}}$ parameter &
\multicolumn{2}{c}{$\mathcal{M}_{_\Sigma}/\bar{\mathcal{M}}_{_\Sigma}(z)$ }\\
& && & & $z=-1$&$z=20$\\
\hline
AFF&$0.27$&$1-\Omega$&$\alpha=0.02333$&&$0.696$&0.692\\
&$0.27$&$1-\Omega$&$\alpha=-0.01667$&&$1.245$&1.248\\
CHA&&&$A=0.7$&$\gamma=0.25$&$0.778$&$0.783$\\
&&&$A=0.75$&$\gamma=-0.1$&$1.171$&$1.166$\\
IDE&&$0.73$&$\xi=2.3$&$w=-1$&$1.306$&$1.297$\\
&&$0.73$&$\xi=3.3$&$w=-1$&$0.9192$&$0.9211$\\
NADE&&&$n=2.5$&&$0.859$ \cite{NADEnote}&$0.864$\\
&&&$n=3.1$&&$1.1125$ \cite{NADEnote}&$1.1118$\\
\end{tabular}
\end{ruledtabular}
\end{table*}
%
\subsection{Flat interacting dark energy}
\begin{figure}[ht!]
\epsfig{file=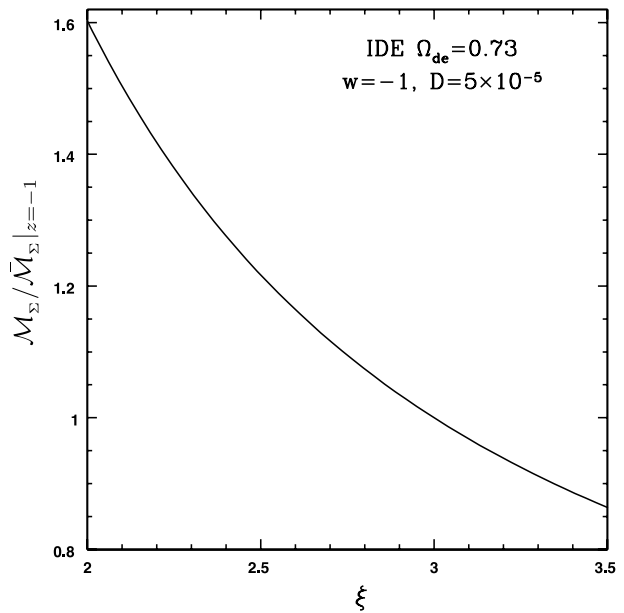,height=3.5in,angle=0}
\caption{Ratio $\mathcal{M}_{_{\Sigma}}/\bar{\mathcal{M}}_{_{\Sigma}}$ evaluated at $z=-1$ as a function of $\xi$ for $w=-1$, $\mathcal{D}=5\times10^{-5}$ and $\Omega_{de}=0.73$ for the interacting dark energy (IDE) model.}\label{IDEzm1}
\end{figure}

It is natural to consider the coupling between dark energy and matter and there are many explicit coupling procedures considered in the literature. Here we use the parameterization of \cite{Dalal01} to write
\begin{equation}
\begin{split}
\left(\frac{H(z)}{H_0}\right)^2=
 (1+z)^3 \left[1-\Omega_{de}\left(1-\frac{1}{(1+z)^\xi}\right)\right]^{-3\left(w/\xi\right)}.
\end{split}
\end{equation}
Figure~\ref{IDEzm1} shows the mass ratio (\ref{massratio}) for the interacting dark energy (IDE) model with $w=-1$, $\Omega_{de}=0.73$ and $\mathcal{D}= 5\times10^{-5}$ at $z=-1$ over a range of $\xi$ consistent with \cite{Dalal01}. The mass ratio (\ref{massratio}) does not  change significantly over the redshift range $-1<z<20$ as can be seen in Table~\ref{table1}.
%
\subsection{Cardassian models}
\begin{figure}[ht!]
\epsfig{file=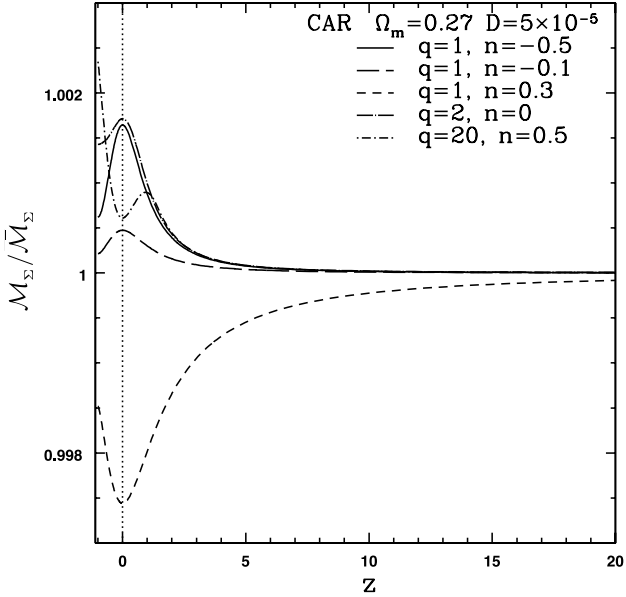,height=3.5in,angle=0}
\caption{Same as Fig.~\ref{CES2} but now $\mathcal{M}$ is given for the cardassian (CAR) model.}\label{CAR}
\end{figure}

The modified polytropic Cardassian models (hereafter CAR) are three parameter models that modify the Friedmann equation in a flat, matter-dominated universe in order to allow acceleration. The Hubble function is given by (see \cite{Davis07,Wang03})
\begin{equation}
\left(\frac{H(z)}{H_0}\right)^2=\Omega_m(1+z)^3\left(1+(\Omega^{-q}_m-1)(1+z)^{3q(n-1)}\right)^{1/q}.
\end{equation}
The evolution of the mass ratio (\ref{massratio}) with redshift for the Cardassian models is shown in Fig.~\ref{CAR} for $\Omega_m=0.27$ and $\mathcal{D}=5\times10^{-5}$ and values of the parameters $n$ and $q$ surrounding the 99.9\% confidence level of \cite{Davis07,Wang03}.
%
\subsection{Flat Chaplygin gas}
\begin{figure}[ht!]
\epsfig{file=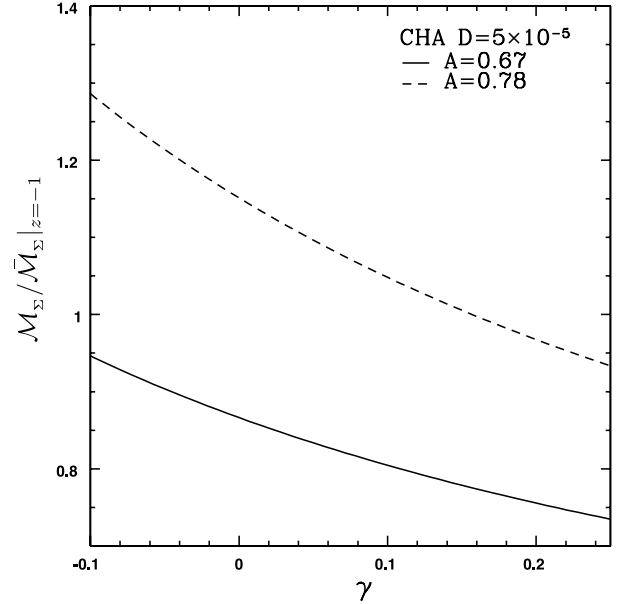,height=3.5in,angle=0}
\caption{Same as Fig.~\ref{IDEzm1} but with $A=0.67$ and $0.78$ as a function of the parameter $\gamma$ for the Chaplygin (CHA) gas model.}\label{CHAzm1}
\end{figure}

A dark fluid that combines dark matter and dark energy, where the equation of state is $p=-A/\rho^\gamma$ \cite{Kamen01}, gives the generalized Chaplygin gas equation (CHA)
\begin{equation}
\begin{split}
\left(\frac{H(z)}{H_0}\right)^2=\left(A+(1-A)(1+z)^{3(1+\gamma)}\right)^{1/(1+\gamma)}.
\end{split}
\end{equation}
Since the Chaplygin gas is a two parameter model we choose two extreme values for $A$ for Fig.~\ref{CHAzm1} in which we plot the mass-ratio (\ref{massratio}) at $z=-1$ as a function of $\gamma$. The values chosen here follow the 99.9\% confidence level of \cite{Davis07}. As with the IDE model, the mass ratio for the Chaplygin gas model is nearly constant over the range $-1<z<20$. See Table~\ref{table1}.
%
\subsection{Flat affine equation of state}
\begin{figure}[ht!]
\epsfig{file=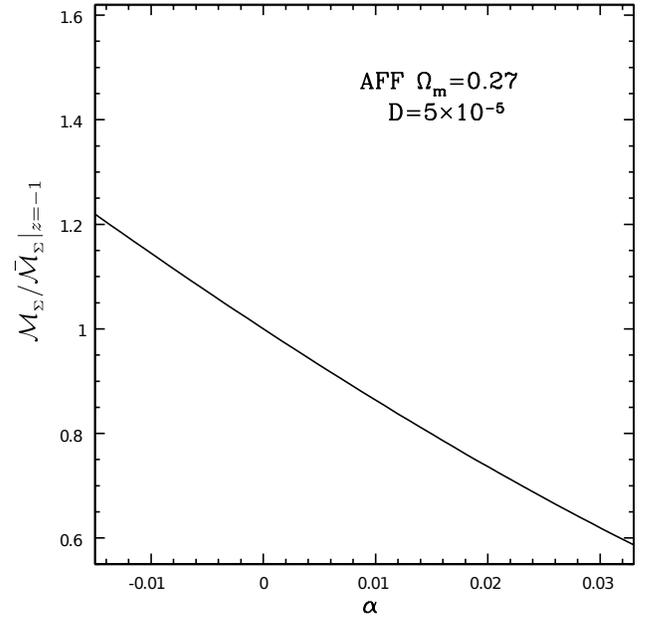,height=3.5in,angle=0}
\caption{Same as Fig.~\ref{IDEzm1} but for the affine (AFF) parameter model as a function of $\alpha$.}\label{AFFzm1}
\end{figure}

The assumption that dark energy and dark matter are a single dark component that can be modeled by the affine equation of state $p=p_0+\alpha\rho$ (AFF) gives rise to the Hubble function \cite{Balbi07}
\begin{equation}
\left(\frac{H(z)}{H_0}\right)^2=\tilde{\Omega}_m(1+z)^{3(1+\alpha)}+\Omega_\Lambda\,,
\end{equation}
where $\tilde{\Omega}_m\equiv(\rho_0-\rho_\Lambda)/\rho_c$.
The ratio (\ref{massratio}) for the AFF model is shown in Fig.~\ref{AFFzm1}  at $z=-1$ as a function of $\alpha$ using the 99\% confidence level of \cite{Balbi07} for $\alpha$. Complementary information is available in Table~\ref{table1}.
%
\subsection{New agegraphic dark energy (NADE)}
\begin{figure}[ht!]
\epsfig{file=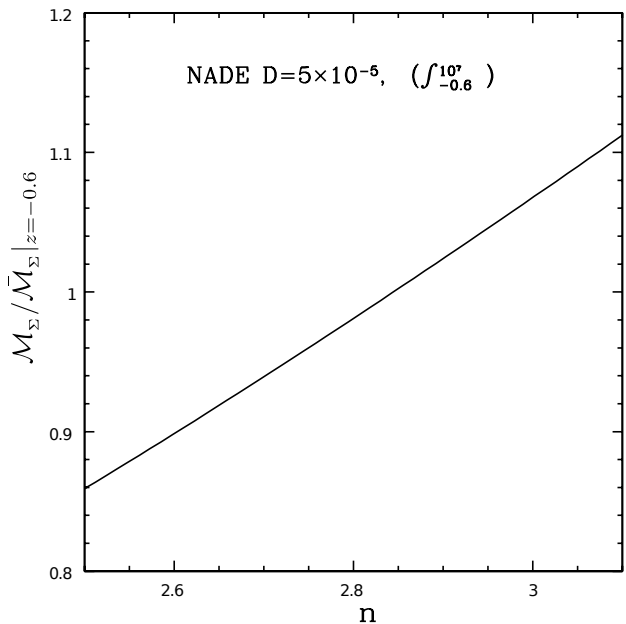,height=3.5in,angle=0}
\caption{Same as Fig.~\ref{CES2}; $\mathcal{M}$ is given for the new agegraphic dark energy (NADE) model \cite{NADEnote}.}\label{NADEzm1}
\end{figure}

A single-parameter model where the energy density of quantum fluctuations of Minkowski spacetime $\rho_q$ is included in the Hubble equation gives \cite{Wei08}
\begin{equation}
\left(\frac{H(z)}{H_0}\right)^2=\sqrt{\frac{\Omega_{m_0}(1+z)^3}{1-\Omega_q(z)}}\,,
\end{equation}
where $\Omega_{m_0}=1-\Omega_q(0)$ and the evolution of $\Omega_q(z)$ is given by
\begin{equation}
\frac{d\Omega_q}{dz}=-\Omega_q(1-\Omega_q)\left[3(1+z)^{-1}-\frac{2}{n}\sqrt{\Omega_q}\right].
\end{equation}
The mass ratio (\ref{massratio}) for this model is presented in Fig.~\ref{NADEzm1} with values of $n$ consistent with the likelihood values in \cite{Wei08}.
%
\section{Discussion}
A property of the mass ratio (\ref{massratio}) seen in all relevant figures is simply
\begin{equation}\label{mproptoh}
\bar{H}(z) \gtrless H(z) \Leftrightarrow \mathcal{M}_{_{\Sigma}} \gtrless \bar{\mathcal{M}}_{_{\Sigma}}.
\end{equation}\\
As can be seen in Fig.~\ref{CES}, \ref{CES2} and \ref{CAR}, the mass ratio (\ref{massratio}) appears to have an extremum at $z=0$. (This feature is present in all plots of the mass ratio as a function of $z$ but is not visible in all figures because of the chosen scale.) In fact, the extremum can not happen at $z=0$. From (\ref{rsigma}) and (\ref{massratio}), and defining $'=d/dz$, we find
\begin{equation}
\begin{split}
\left(\frac{\mathcal{M}_{_\Sigma}}{\bar{\mathcal{M}}_{\Sigma}}\right)'&=3\frac{r_{_\Sigma}^2}{\bar{r}_{_\Sigma}^4}\left\{r_{_\Sigma}'\bar{r}_{_\Sigma}-\bar{r}_{_\Sigma}'r_{_\Sigma}\right\}\,,
\end{split}\label{Mprime}
\end{equation}
where from (\ref{rsigma})
\begin{equation}\label{rprime}
r'_{_\Sigma}(z)=- \sqrt{\frac{\mathcal{D}^2(1+z)^2}{1+\mathcal{D}^2(1+z)^2}}\frac{1}{H(z)}
\end{equation}
and equivalently for the $\Lambda$CDM model (with a bar). Therefore the mass ratio is extremal for
\begin{equation}\label{extremal}
    \frac{r_{_\Sigma}}{\bar{r}_{_\Sigma}}=\frac{r'_{_\Sigma}}{\bar{r}'_{_\Sigma}}=\frac{\bar{H}}{H}=\left(\frac{\mathcal{M}_{_\Sigma}}{\bar{\mathcal{M}}_{\Sigma}}\right).
\end{equation}
An extremum at $z=0$ would imply that $r_{_\Sigma}=\bar{r}_{_\Sigma}$ and that $\mathcal{M}_{_\Sigma}=\bar{\mathcal{M}}_{_\Sigma}$. Equation (\ref{Mprime}) can also be expressed as
\begin{equation}\label{Mprime2}
\left(\frac{\mathcal{M}_{_{\Sigma}} }{\bar{\mathcal{M}}_{_{\Sigma}}}\right)'_{_{z=0}}=3\delta^3_{0}\sqrt{\frac{\mathcal{D}^2}{1+\mathcal{D}^2}}\frac{1}{r_{_{\Sigma}}(0)H_0}\{\delta_0-1\}\,,
\end{equation}
where $\delta_0^3=\left(\mathcal{M}_{_{\Sigma}}(0)/\bar{\mathcal{M}}_{_{\Sigma}}(0)\right)$ and $r_{_{\Sigma}}(0)$ is obtained from (\ref{rsigma}) for the model investigated. 

In Eq.~($\ref{rsigma}$), the influence of $\mathcal{D}$ in the first factor of the integral converges to $1$ for $\mathcal{D}^2(1+z)^2\gg1$ or $1+z\gg1/\mathcal{D}$ and converges to $\mathcal{D}(1+z)$ when $1+z\ll1/\mathcal{D}$. In effect then since $\mathcal{D}\sim10^{-5}$, $\mathcal{D}$ is unimportant at early times for ${r_{_\Sigma}}$ and $\bar{r}_{_\Sigma}$.
%
\section{CONCLUSION}
We have introduced a new generalized Swiss-cheese model which does not assume \emph{a priori} that the associated boundary surfaces are comoving. In order to quantify evolving inhomogeneities, we have considered geodesic boundaries characterized by their linear momentum $\mathcal{D}$. For the size of the inhomogeneities we are interested in, the physical values of $\mathcal{D}/c$ are $\sim10^{-5}$. For a given linear momentum, we have found that the inhomogeneities grow almost independently of the background model (with the inclusion of the radiation density parameter $\Omega_{R_0}$). As shown in Fig.\ref{rsigmabar} these inhomogeneities are almost at their full size by the decoupling ($z\sim 1100$). For a redshift of $z \lesssim 2$, corresponding to high redshift supernovae, the inhomogeneities considered here are growing very slowly as is shown in see Fig.\ref{MSMLCDM}.
\section*{ACKNOWLEDGMENTS}
CG is supported by the Ontario Graduate Scholarship in Science and Technology and the Fonds qu\'eb\'ecois de recherche sur la nature et les technologies. KL is supported by a grant from the Natural Sciences and Engineering Research Council of Canada. Portions of this work were made possible by use of \emph{GRTensorII}\cite{GRTensorII}.
\section*{APPENDIX: RADIATION}\label{radiation}
The integration in Eq.~(\ref{rsigma}) requires a high redshift contribution from radiation. See, for example, \cite{Peebles93,Komatsu09}. This was added to all the Hubble background models $H(z)$ as noninteracting species $\propto a^{-4}$. The energy density for radiation $\rho_R$ is
\begin{equation}
\begin{split}
\rho_R&=a_B T_{\rm CMB}^4\left[1+\frac{7}{8}\left(\frac{T_\nu}{T_{\rm CMB}}\right)^4N_\nu\right]\,,
\end{split}\label{rhorad}
\end{equation}
where $a_B=4\sigma/c$ in which $\sigma$ is the Stefan-Boltzmann's constant, $T_{\rm CMB}=2.725K$ is the temperature of the cosmic microwave background (CMB), $T_\nu=(4/11)^{1/3}T_{\rm CMB}$ is the temperature parameter for the relativistic neutrinos after the annihilation of electron-positron pairs and $N_\nu=3$ is the standard number of neutrino families. The first right hand side term is the contribution from the CMB photons and the second term is the contribution from relativistic neutrinos. The energy density of radiation is therefore:
\begin{equation}\label{densrad}
\Omega_{R_0}=\frac{\rho_{R_0}}{\rho_{c_0}}\simeq8\times10^{-5}
\end{equation}
where $\rho_c$ is the critical energy density; the subscript $0$ is used to specify present values ($z=0$).
Note that before the electron-positron pairs annihilated ($z\sim10^{10}$) the temperature of the neutrinos and CMB radiation was in equilibrium $T_\nu=T_{\rm CMB}$. This contribution in Eq.~(\ref{rhorad}) was negligible in our integration of (\ref{rsigma}) and was ignored.



\begin{thebibliography}{}
\bibitem{notation}We use geometrical units, a signature of $+2$ and designate functional dependence usually only on the first appearance of a function. When the velocity of light enters explicitly, we designate it by $c$.
\bibitem{covariant} The noncovariant character of (\ref{mass}) is clarified in \cite{pe,z,h} essentially as follows: Consider the three spacelike Killing vector fields $\eta_{(i)}$ obeying the algebra of SO(3) ($[\eta_{(i)},\eta_{(j)}]=\epsilon_{ijk}\eta_{(k)}$) and define $L^2=\frac{1}{2}\Sigma_{i}\eta^{\alpha}_{(i)} \eta_{(i) \alpha}$. Then $\mathcal{M}=\frac{L}{2}(1-\nabla^{\alpha}L\nabla_{\alpha}L$).
\bibitem{lake1}Whereas this equation is derived in \cite{lake}, the physical meaning of $\mathcal{D}$ was not given there.
\bibitem{NADEnote} We could not integrate down to $z=-1$ nor up to $z=\infty$ with the NADE model using {\sc maple} or {\sc mathematica}. Our reliable integration limits are $z=-0.6$ and $z=10^7$.
\end{thebibliography}
\end{document}